\documentclass[preprint]{iucr}              % DO NOT DELETE THIS LINE
\usepackage{bm}
\usepackage{amsmath}
\usepackage{amssymb}
\usepackage{graphicx}
\usepackage{array}
\usepackage{makeidx}
\usepackage{siunitx}

     \journalcode{S}              % Indicate the journal to which submitted
\begin{document}                  % DO NOT DELETE THIS LINE

\title{A pipeline for Megahertz X-ray Photon Correlation Spectroscopy on soft matter samples at the MID instrument of European XFEL}

\cauthor[a, b]{Aliaksandr}{Leonau}{aliaksandr.leonau@xfel.eu}

\author[b]{Felix}{Brausse}
\author[b]{James}{Wrigley}
\author[b]{Mads B.}{Jakobsen}
\author[a]{Amir}{Tosson}
\author[a]{Michelle}{Dargasz}
\author[a]{Nimmi Das}{Anthuparambil}
\author[c, d]{Felix}{Lehmkühler}
\author[e]{Anita}{Girelli}
\author[e]{Maddalena}{Bin}
\author[e]{Fivos}{Perakis}
\author[f]{Sebastian}{Retzbach}
\author[f]{Fajun}{Zhang}
\author[f]{Frank}{Schreiber}
\author[b]{Matheus}{Teodoro}
\author[b]{Cammille}{Carinan}
\author[b]{Robert}{Rosca}
\author[b]{Fabio}{Dall'Antonia}
\author[b]{Wonhyuk}{Jo}
\author[b]{Ulrike}{Boesenberg}
\author[b]{Angel}{Rodriguez-Fernandez}
\author[b]{Roman}{Shayduk}
\author[b]{Jörg}{Hallmann}
\author[b]{Alexey}{Zozulya}
\author[b]{Jan-Etienne}{Pudell}
\author[b]{Carsten}{Deiter}
\author[b]{Luca}{Gelisio}
\cauthor[b]{Johannes}{Möller}{johannes.moeller@xfel.eu}

\author[b]{Anders}{Madsen}
\author[a]{Christian}{Gutt}

\aff[a]{Department Physik, Universität Siegen, Walter-Flex-Strasse 3, 57072 Siegen, \country{Germany}}

\aff[b]{European X-Ray Free-Electron Laser Facility, Holzkoppel 4, 22869 Schenefeld, \country{Germany}}
\aff[c]{Deutsches Elektronen-Synchrotron DESY, Notkestr. 85, 22607 Hamburg, \country{Germany}}
\aff[d]{The Hamburg Centre for Ultrafast Imaging, Luruper Chaussee 149, 22761 Hamburg, \country{Germany}}
\aff[e]{Department of Physics, AlbaNova University Center, Stockholm University, S-106 91 Stockholm, \country{Sweden}}
\aff[f]{Institut für Angewandte Physik, Universität Tübingen, Auf der Morgenstelle 10, 72076 Tübingen, \country{Germany}}

\maketitle                        % DO NOT DELETE THIS LINE

\begin{abstract}
In this paper we present the experimental protocol and data processing framework for Megahertz X-ray Photon Correlation Spectroscopy (MHz-XPCS) experiments on soft matter samples, implemented at the Materials Imaging and Dynamics (MID) instrument of the European X-ray Free-Electron Laser (EuXFEL). Due to the introduction of a standard configuration and the implementation of a highly automated data processing pipeline, MHz-XPCS measurements can now be conducted and analyzed with minimal user intervention. A key challenge lies in managing the extremely large data volumes generated by the Adaptive Gain Integrating Pixel Detector (AGIPD) - often reaching several petabytes within a single experiment. We describe the technical implementation, discuss the hardware requirements related to effective parallel data processing, and propose strategies to enhance data quality, in particular related to data reduction strategies and an improvement of the signal-to-noise ratio. Finally, we address strategies for making the processed data FAIR (Findable, Accessible, Interoperable, Reusable), in alignment with the goals of the DAPHNE4NFDI project. 

\end{abstract}

\section{Introduction}

The development of hard X-ray sources with improved coherence properties, such as X-ray Free-Electron Lasers (XFELs) and diffraction-limited storage rings, makes X-ray photon correlation spectroscopy (XPCS) a powerful tool for studying dynamics in hard and soft condensed matter \cite{gruebel2007, zhang2018, perakis2020, lehmkuhler2021}. These sources provide beams of high spatial coherence allowing detection of speckle fluctuations in the scattered intensity to reveal nanoscale dynamics of proteins, polymers, and other biological or soft matter structures. As these advanced X-ray sources become more prevalent and accessible, XPCS is increasingly being exploited in the study of dynamic processes in soft matter and biological systems \cite{moeller2019,begam2021,girelli2021,reiser2022,anthuparambil2023, dallari2024, otto2024}. This development opens new avenues for routine investigation of complex, time-dependent phenomena in biology and soft matter science.

This progress is in some ways equivalent to the development of Biological Small-Angle X-ray Scattering (BioSAXS), which evolved from the early conception as a specialized method into a routine tool in structural biology with the emergence of third-generation synchrotron radiation sources \cite{Blanchet2013, grawert2020}. The widespread adoption of BioSAXS was thrusted by the development of dedicated experimental setups \cite{classen2013, acerbo2015, martel2012, pernot2013, thureau2021, cowieson2020, blanchet2015, fujisawa2000}, data analysis tools \cite{forster2010, manalastas2021, hopkins2024} and a common database for experimental data \cite{kikhney2020}, facilitating non-expert users to obtain reproducible and high-quality results. Similarly, the establishment of standardized protocols and data analysis tools will be crucial in making XPCS accessible for a broader scientific community, for instance to contribute with a better understanding of the role of dynamics in phase transitions and cellular functions.

The European X-ray Free-Electron Laser facility (EuXFEL) is the first XFEL to produce ultrashort hard X-ray pulses in bursts of megahertz (MHz) repetition rate \cite{decking2020}. This high repetition rate, combined with the exceptional (laser-like) spatial coherence properties, enable MHz X-ray Photon Correlation Spectroscopy (MHz-XPCS) to probe diffusive dynamics with (sub-) microsecond temporal resolution \cite{Lehmkuehler2020PNAS,dallari2021interactions,lehmkuhler2021, madsen2021}. The resulting time and length scales match typical diffusion processes in dense cellular environments, therefore enabling the study of complex many-body interactions between proteins and their solvent at molecular length scales \cite{reiser2022, girelli2024, anthuparambil2025}. This capability allows MHz-XPCS to access a range of length and time scales that optical techniques such as dynamic light scattering (probing larger length scales) and neutron spectroscopy techniques (measuring faster time scales) cannot reach.

However, the high repetition rate in combination with the use of pixelated, two-dimensional detectors to record the data, results in very large data volumes. This requires advanced data handling solutions, including high-performance computing and efficient data storage systems \cite{malka2023}, as well as specialized analysis software to effectively manage and quickly calculate the desired correlation functions. In this work we outline the implementation of an experimental configuration as well as a data analysis pipeline \cite{extra_speckle} that was customized to the burst-mode acquisition scheme of EuXFEL. The described experimental setup and tools are available for all users of the Materials Imaging and Dynamics (MID) instrument of EuXFEL \cite{madsen2021}.

\section{Experimental setup and measurement protocol}

The MID instrument of EuXFEL is used in the standard SAXS configuration, described in detail in \cite{madsen2021}. X-rays are generated by the SASE-2 undulator with a typical photon energy of $\sim 10$ keV. The stochastic Self-Amplified Spontaneous Emission (SASE) process results in a pulse bandwidth of $\sim 30$\,eV. Narrower bandwidth radiation, with correspondingly improved longitudinal coherence, is optionally available via hard X-ray self-seeding \cite{liu2023,Boesenberg2024} or by use of crystal monochromators \cite{Petrov2022, Tasca2023, zozulya2025}.

\begin{figure}
    \begin{center}
        \includegraphics[width=1\textwidth]{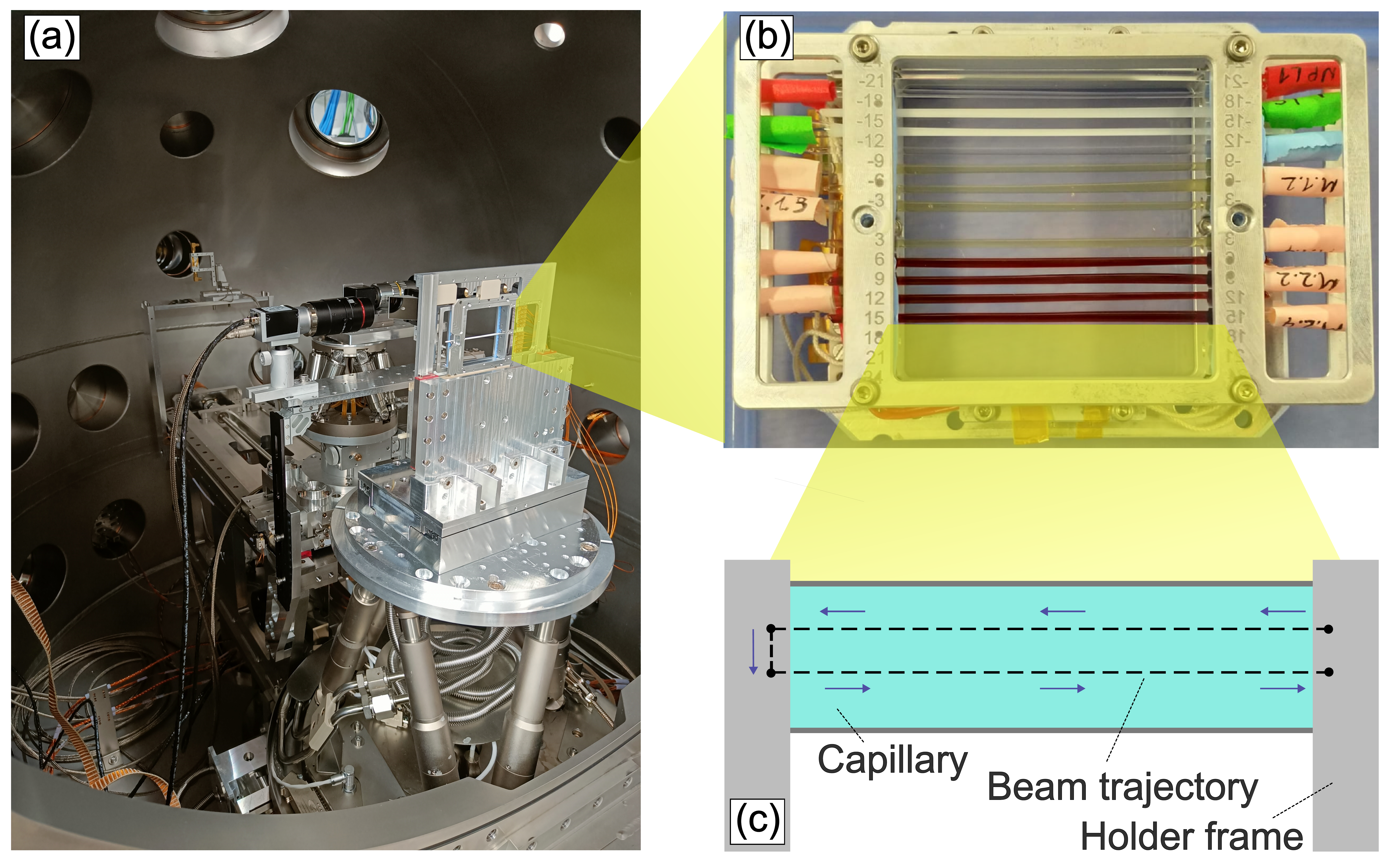}
        \caption{(a) Scanner stage and (b) holder for 15 capillaries used at the MID instrument. (c) Schematic of the beam trajectory (dashed lines) during the scan of a single capillary in one measurement run.}
        \label{fig:stage_holder}
    \end{center}
\end{figure}

The X-ray beam is collimated and subsequently focused onto the sample positions using two transfocator untis equipped with beryllium compound refractive lenses (CRLs) at locations of $229$\,m and $931$\,m from the source. The sample position is located $959$\,m from the source. The sample chamber hosts different stages for sample manipulation, including a hexapod (H-840, Physik Instrumente) and a fast scanner stage (see Fig. \ref{fig:stage_holder}a). This scanner stage features an exchangeable frame system, which can be loaded via a loadlock system without venting the sample chamber. Standardized frames, which can hold up to 15 glass capillaries are available for users (Fig. \ref{fig:stage_holder}b), but customized solutions are also possible. Measuring samples in air is possible as well, enabled by the use of two diamond windows upstream and downstream of the interaction region \cite{madsen2021}.

XPCS experiments, especially in SAXS geometry on soft matter samples, are often limited not by the coherent flux, but rather by beam induced effects that influence the sample's structure and/or dynamics. Beam induced dynamics have been observed even for hard condensed matter samples \cite{ruta2017, pintori2019}, however, for absorbed doses in the range of megagray (MGy), and beyond. Protein samples in solution can typically tolerate a dose up to $1 - 10$ kGy \cite{jeffries2015} at dose rates not exceeding $\sim 1$~kGy~$\mu$s$^{-1}$ in a time window up to several tens of microseconds \cite{reiser2022, Timmermann2023}. Therefore, the experimental task is to optimize the signal-to-noise ratio (SNR), while minimizing the dose and dose rate on the sample. Experiments are performed with only modestly focused X-ray beams, which, in return, requires large sample detector distances to retain sufficient speckle contrast despite the smaller speckle size. In the case of MID, this is realized by a beam size of $\sim 15$~$\mu$m and by placing the detector at $\sim 7$~m downstream of the sample. At a typical photon energy of $\sim 10$\,keV, this results in a relatively low speckle contrast of only a few percent, since the speckle size of about $58$\,$\mu$m is much smaller than the pixel size of the Adaptive Gain Integrating Pixel Detector (AGIPD) ($200$\,$\mu$m) \cite{Allahgholi2019,Sztuk2023}. Next-generation detectors with smaller pixel size would improve the situation, possibly matching the pixel and speckle size for optimal SNR \cite{lumma2000, moeller2019, madsen2020structural}. Experiments with significantly larger sample detector distance are not possible due to space constraints and the reduced scattering angle range covered by the detector, and a smaller X-ray focal spot size is also not favored due to beam heating/damage effects \cite{Lehmkuehler2020PNAS,Xu:22,dallari2024,tanner2025}.

Another means to optimize the SNR while enforcing the radiation dose constraints is spreading the measurements over a large  scattering volume (many sample positions), with only a limited number of frames acquired at each position \cite{moeller2019, reiser2022}. This is enabled by the burst mode operation of EuXFEL \cite{decking2020}, which delivers a short train of X-ray pulses (up to a few hundred) every $100$\,ms. A modest translation speed of the sample, typically $0.4$\,mm/s, results in the separation by multiple beam spot sizes between two consecutive trains, while the translation is negligible ($<100$\,nm) within a train. Therefore, an independent correlation function is measured with each train at a fresh spot on the sample (Fig. \ref{fig:stage_holder}c). This acquisition scheme not only distributes the radiation dose evenly on the sample, but also allows the application of a necessary data correction scheme as detailed in the next sections.

\section{Data processing pipeline}

\begin{figure}
    \begin{center}
        \includegraphics[width=1\textwidth]{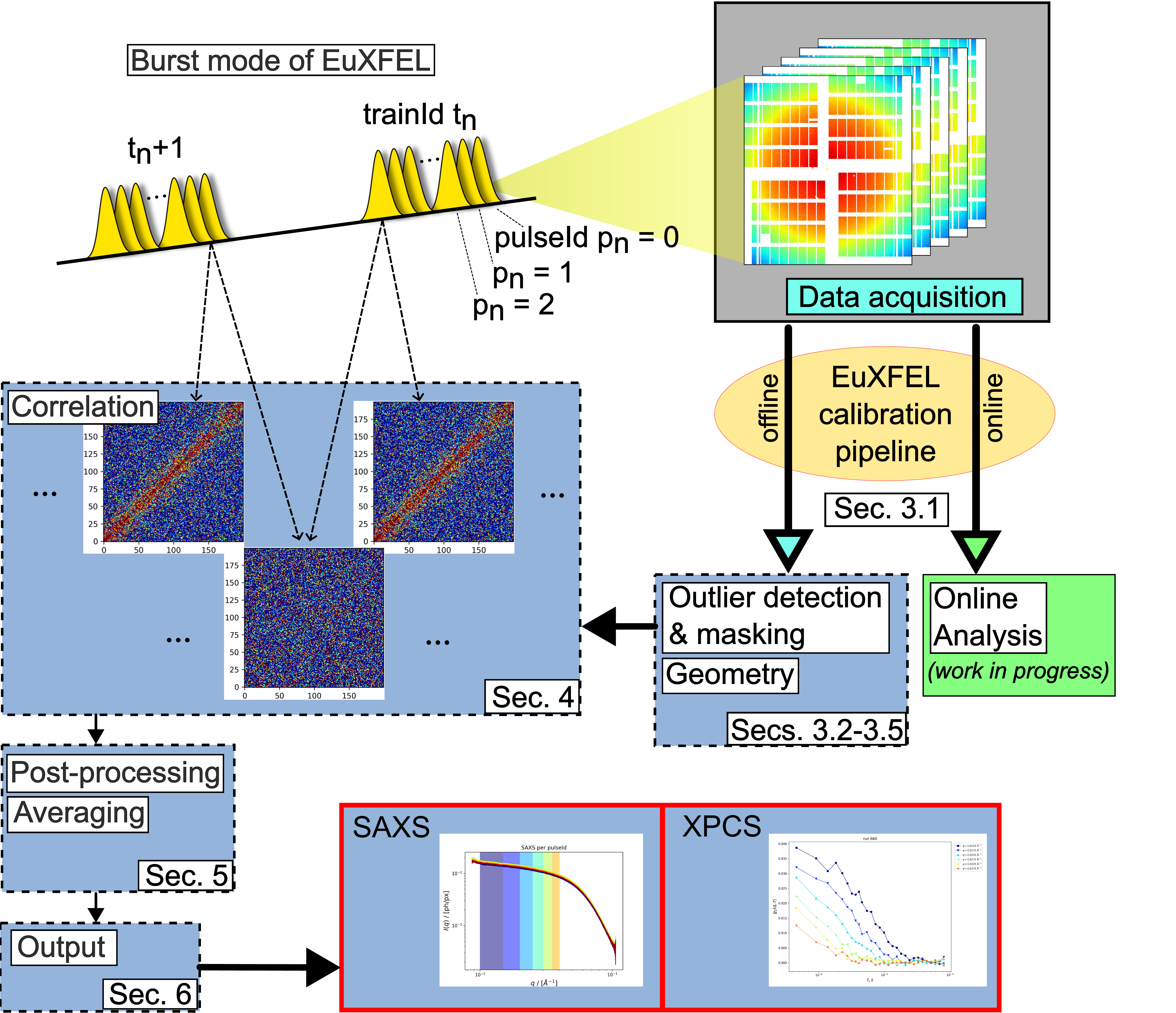}
        \caption{Sketch of the data acquisition and processing workflow for MHz-XPCS at the European XFEL. Scattering patterns are recorded by the AGIPD at acquisition rates matching the EuXFEL time structure (X-ray pulses spaced by several hundred nanoseconds are delivered in pulse trains at a 10 Hz repetition rate). Each frame is labeled with a unique combination of pulse and train Id. The raw detector data is processed by the dedicated EuXFEL calibration pipeline which produces corrected datasets (''proc'' data). Once the processed detector data become available on the offline cluster, the MHz-XPCS pipeline initiates further processing steps, including geometry and direct beam refinement, outlier detection, additional filtering, region-of-interest (ROI) specification, etc. The train-resolved two-time correlation functions and off-correlations are then calculated. Following several post-processing steps (additional threshold filtering, averaging, etc.) the final output is generated and made accessible to the users. An online XPCS pipeline - currently under development - receives and processes data streamed directly from the online calibration pipeline and will be described in a separate publication.}
    \end{center}
    \label{fig:geom}
\end{figure}

A schematic overview of the subsequent data handling and processing steps is shown in Fig. \ref{fig:geom} using representative scattering data  from ferritin in solution for illustration. The different steps indicated in the figure will be described in detail in the following sections. 

\subsection{Data acquisition and EuXFEL calibration pipeline}

The coherent small-angle X-ray scattering data are collected using the AGIPD in high-CDS mode \cite{Allahgholi2019,Sztuk2023}. In this configuration, the adaptive gain feature of the detector is not supported, but the mode offers a slightly better sensitivity for single photon detection, which helps to mitigate certain detector artifacts, as explained in later sections. The detector comprises $n_{mod} = 16$ individual modules, each having a dimension of $n_{pix} = 512 \times 128$ pixels.

The AGIPD was specifically developed to capitalize on the opportunities provided by EuXFEL's burst mode and is capable of  recording data at frame rates of up to $4.5$\,MHz. During each pulse train, up to $n_p = 352$ pixel values can be stored at this frame rate and subsequently read-out 10 times per second corresponding to the repetition rate of trains (burst mode). For a typical recorded run (the term used at EuXFEL to describe an act of data acquisition, analogous to the meaning of scan at storage rings) containing $n_{tr}\sim 2600$ trains, this results in an overall volume of $n_{tr} \times n_p \times n_{mod} \times n_{pix} \times 2~\mathrm{bytes} \times 2 \approx 3.8~\mathrm{TB}$ for $\sim 4.5$ minutes of data collection. Although the last factor of 2 is due to simultaneous recording of intensity and gain stage information, which can be omitted here, this data rate still generates several petabyte (PB) of data  over the course of a few days of experiment \cite{Sztuk2023,sobolev2024}. 

To convert the raw detector output into corrected data suitable for analysis, two calibration pipelines have been developed at EuXFEL, with the main concepts and technical details described in \cite{Schmidt2024, sobolev2024,Sztuk2023}. A distinction is made between online and offline analysis, which are implemented in parallel. Details on the general data analysis software and infrastructure at EuXFEL is available in \cite{dadocumentation}. The online pipeline corrects and streams part of the detector data to dedicated analysis nodes, enabling  immediate data analysis and visualization. This part of the analysis aims at minimum latency in order to guide the running experiment, but with potentially only a subset of the overall data being analyzed. An XPCS analysis pipeline based on online data is currently being developed and will be described in a later publication.

The offline pipeline \cite{extra_speckle} is based on the input from  EuXFEL's HDF5 data files recorded, stored, and transferred to the DESY computing infrastructure (Maxwell cluster) \cite{MaxwellHPC} prior before being calibrated. Here, the objective is to process the full dataset with highest possible quality, accepting higher latency. This data is further processed by the XPCS pipeline (dashed frames in Fig.~\ref{fig:geom}) described in the present paper.

Correction procedures specific to the AGIPD have been developed and refined in the early years of EuXFEL operation \cite{Sztuk2023}. While high dynamic range and gain switching are essential for some applications, the quality of XPCS data depends on the noise characteristics and single-photon sensitivity. To this end, flat field corrections based on Cu fluorescence data, common-mode corrections, and conversion to absolute photon numbers ("photonizing") have been shown to significantly improve the quality of the XPCS results  \cite{dallari2021analysis}. These correction steps are now integrated into the standard AGIPD calibration pipeline used in user experiments \cite{Sztuk2023}. Additionally, pixels are filtered based on a set of criteria, from which a so-called bad pixel mask (calibration mask) is generated. The corrected data is termed "proc" and saved in designated HDF5 files. Due to photonizing (integers) and the sparse nature of XPCS data, these files exhibit compression ratios of $\sim 50$ ($\sim 25$ when the calibration mask is included) \cite{sobolev2024}.

While these correction steps already improve the data quality significantly, certain detector artifacts remain visible in the two-time correlation functions (TTCFs) \cite{madsen2010}. These artifacts are linked to drifts and jumps of the pedestal, which are not always caught by the common mode correction (see Appendix for details). Due to the storage cell and pixel read-out architecture of the AGIPD such "jumping" pixels frequently affect blocks of 32 storage cells simultaneously resulting in square-like features in the TTCF (an example is shown in Fig.~\ref{fig:ttcf_off}). The random occurrence of these jumps cannot be assigned to a few damaged pixels, suggesting they are an intrinsic feature of the AGIPD. Therefore, special attention must be given to identifying and correcting these "jumping" pixels in each recorded dataset. Implementing these additional correction steps imposes a computational cost: the data set must be loaded twice for outlier detection prior to correlation calculations (Sec. \ref{sec:outlier_removal}) - and the number of calculated correlations is effectively doubled due to the off-correlation corrections (Sec. \ref{sec:off_corr}). Hence, in Sec. \ref{sec:future} we discuss possible scenarios to further improve the data quality and reduce the data footprint in the future.
\subsection{Mean intensity calculation}
\label{ch3:mpc}

\begin{figure}
    \begin{center}
        \includegraphics[width=1.0\textwidth]{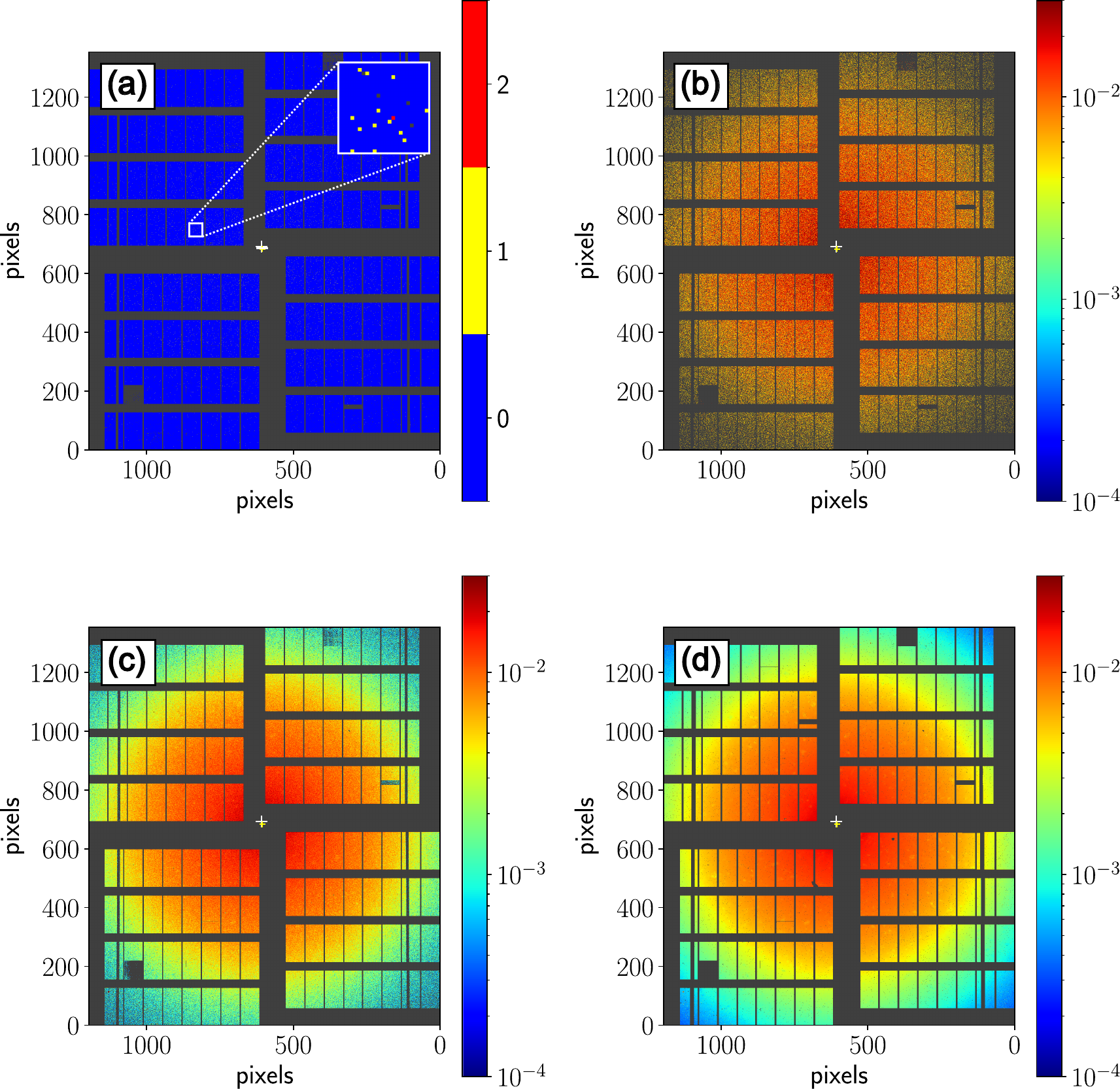}
        \caption{2D scattering patterns of ferritin in solution collected with the AGIPD shown at different levels of averaging. (a) Single pulse; (b) single train averaged over $200$ pulses; (c) average over 10 trains; (d) single run averaged over 2200 trains. The color scale indicates the number of photons per pixel.}
        \label{fig:pls_tr_run}
    \end{center}
\end{figure}

As a first step of the XPCS pipeline, the mean intensity is calculated for each storage cell of every pixel. Figure~\ref{fig:pls_tr_run} illustrates examples of a single acquisition along with subsequent averages over increasing number of frames. This initial step is required to locate detector pixel outliers (see Sec.~\ref{sec:outlier_removal} below) that need to be masked in the subsequent calculation steps and additionally offers the possibility of refining the scattering geometry. Pixels already marked as bad pixel by the standard calibration pipeline are already masked and excluded at this stage. In addition, users can supply a static mask for excluding shaded regions on the detector and apply filtering with respect to the sample scanner position (to remove data collected during illumination of the sample holder frame, see Fig.~\ref{fig:stage_holder}c). %This results in the total reduction $2600\rightarrow 2200$ trains with data to be processed. It is important to stress that starting and stopping the scan as well as shifting onto a new line in the vertical direction is done on the frames in order to avoid radiation damage and even local boiling of the sample. \textcolor{red}{\textbf{[should we tell that it is related to DAQ and shutter, etc.?]}} 

The mean intensity is then calculated as the intensity averaged over trains:
\begin{equation}
    \bar{I}(p_n,m,x,y) = \bigl\langle I(t_n,p_n,m,x,y) \bigr\rangle_{t_n},
\label{eq:mean_pixelcell}
\end{equation}

\noindent with $t_n$ and $p_n$ designating the train-ID and the pulse-ID, respectively; $m =\{ 0; 1; ...; 15\}$ enumerates the detector module; $x$ and $y$ are the pixel coordinates (integers) within each module. Calculating the mean intensity of the pixels using (\ref{eq:mean_pixelcell}) benefits from the structure of the AGIPD data, which is saved for each of the $n_{mod}=16$ modules in separate HDF5 files. Furthermore, the data is divided into groups of 250 trains ("sequences") for each sequence file. For a typical run containing approximately 2600 trains, this results in $\sim 176$ module/sequence files ($n_{mod} \times 11$) which can be processed efficiently in parallel using multiprocessing.

\begin{figure}
    \begin{center}
        \includegraphics[width=0.80\linewidth]{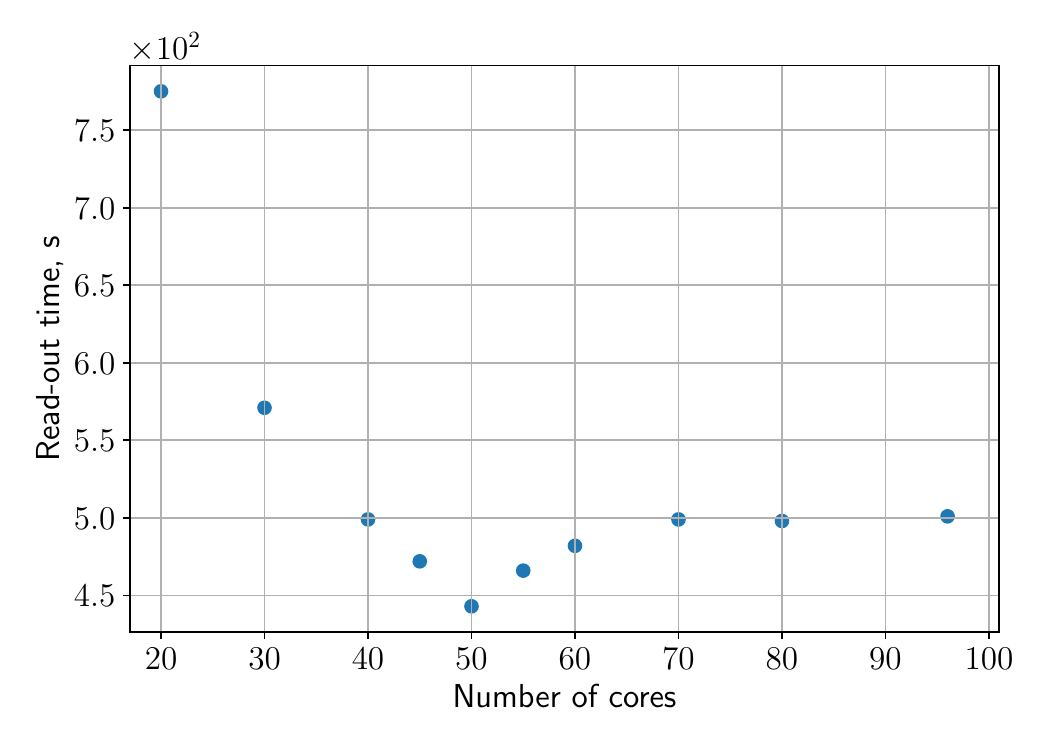}
        \caption{Performance of the parallel read-out of AGIPD data files as a function of the number of CPU cores used on the DESY Maxwell cluster.}
        \label{fig:h5_read}
    \end{center}
\end{figure}

Performing large-scale parallel I/O operations, even on high-performance computing resources such like the  Maxwell cluster, can be demanding and saturating the available bandwidth. Hence, we benchmarked the parallel read-out of data required for calculating Eq. (\ref{eq:mean_pixelcell}) using a varying number of CPU cores. Based on the results shown in Fig.~\ref{fig:h5_read}, we chose a default value of $n_{core}=50$ cores for this task (with the option to override this value in the specific pipeline call). To avoid running out of memory, the data are loaded train-wise by each core and summed within the existing data container. This results in $\sim 20$\,GB of simultaneous memory load when using the default number of cores ($n_{core} \times n_p \times n_{pix}  \times 17\,\mathrm{bytes} = 19.6~\mathrm{GB} $ with the factor 17 corresponding to the size (byte) of containers to load image / mask and save the data).

%In addition to calculating the mean detector image, we also obtain the mean sacttering intensity within each train. These values can be used for further data filtering before executing the XPCS calculation.

\subsection{Scattering geometry}
\label{sec:scatt_geo}

A quantitative evaluation of the X-ray scattering data requires calibration of the momentum transfer (scattering vector) $\vec{q}$, which depends on the exact geometry of the scattering experiment. To support a high level of automatization, we have implemented automatic retrieval routines into the data pipeline.

The 16 separate modules of AGIPD are grouped in 4 quadrants (with 4 modules forming each quadrant). Each quadrant can be translated in both vertical and horizontal directions for flexible coverage of different regions in reciprocal space \cite{allahgholi2019adaptive}. The respective motor positions are recorded alongside the scattering data and are used to calculate the position of each quadrant of the detector using the extra-geom package of EuXFEL \cite{extrageom}.

\begin{figure}
    \begin{center}
        \includegraphics[width=1.0\textwidth]{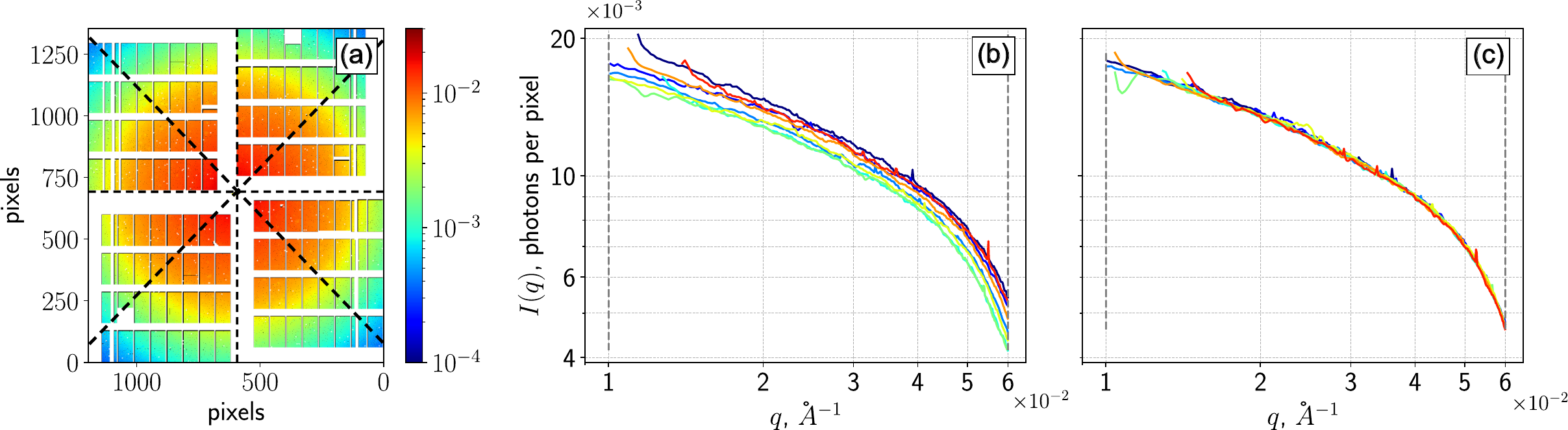}
        \caption{Refinement of the direct beam position applied to ferritin scattering data. (a)~"Pizza slicing" of the mean detector image into 8 equal azimuthal sectors. Azimuthally intergated intensity profiles $I(q)$ for each sector are shown (b)~before and (c)~after optimization. Improved overlap after refinement indicates a more accurate beam center determination. }
        \label{fig:beam_center_ref}
    \end{center}
\end{figure}

To automatically determine the beam center, defined as the position of the incident, directly transmitted beam intersecting the detector plane, we exploit the azimuthal symmetry of the scattering signal and the fact that the mean detector image is already held in memory for the XPCS analysis. An initial estimate of the beam center position can be provided by the user or set by the pipeline to the geometric center of the central rectangular gap between the detector quadrants. Subsequently, an iterative optimization algorithm is applied, which slices the mean detector image into eight equal azimuthal ranges (as shown in Fig. \ref{fig:beam_center_ref}a)  and carries out separate azimuthal integrations of the intensity \cite{ashiotis2015, Kieffer2020}. This procedure is repeated with different direct beam positions and iterated to maximize the overlap among the resulting azimuthally integrated intensity profiles $I(q)$. The search for a direct beam position is performed within the central gap of the AGIPD. An example of the refinement process is shown in Fig. \ref{fig:beam_center_ref}b-c.

\subsection{Outlier pixel detection}
\label{sec:outlier_removal}
In order to mask "jumping" pixels prior to XPCS calculations, outliers are identified by comparing pixels expected to exhibit similar mean intensities and counting statistics. After the automatic determination of the scattering geometry (as described in Sec.~\ref{sec:scatt_geo}), each pixel is compared against others within the same narrow $q$-interval. By default, the AGIPD detector area is split into $300$ concentric annular regions of fixed width but increasing radius, such that all pixels within a given annulus correspond to the same $q$- value. For each pixel, the root mean square (RMS) value is calculated along the pulse-ID dimension as 

\begin{equation}
    RMS\bigl[I(m,x,y)\bigr] = \sqrt{\bigl\langle \bar{I}(p_n,m,x,y)^2 \bigr\rangle_{p_n}}.
\label{eq:rms}
\end{equation}

\begin{figure}
    \begin{center}
        \includegraphics[width=1\textwidth]{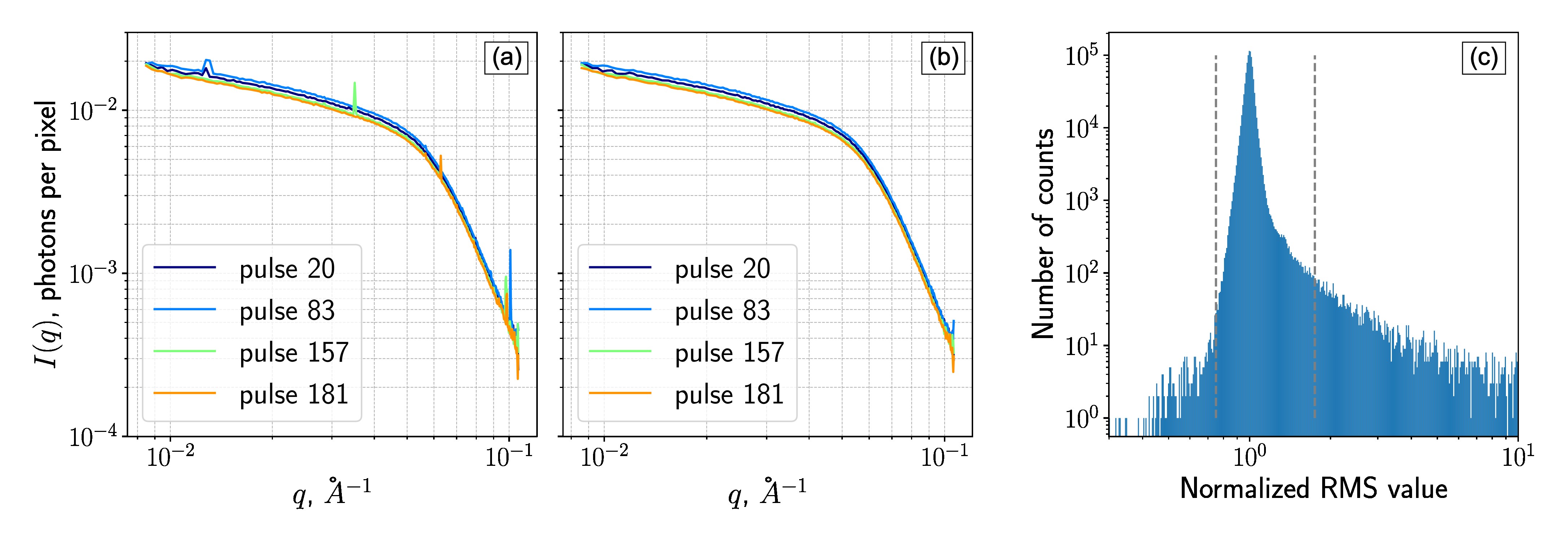}
        \caption{Removing of pixel outliers based on train averaged ferritin scattering data. (a)Azimuthally integrated intensity $I(q)$ for selected pulses before outlier filtering. (b) $I(q)$ for the same pulses after filtering. (c)~Distribution of normalized RMS values used for outlier detection. The dashed lines indicate the lower and upper threshold values $0.75$ and $1.75$, respectively.}
    \end{center}
    \label{fig:out}
\end{figure}

\noindent Subsequently, the resulting RMS value is normalized by the median of all RMS values of pixels within the same $q$ region. A pixel is flagged as an outlier if its normalized RMS-value lies outside the empirically determined window of $0.75$ to $1.75$. These bounds are not symmetric around 1 reflecting the skewed nature of the RMS (see Fig.~\ref{fig:out}c). Both values are user-configurable. 

The effect of outlier removal is demonstrated in Fig.~\ref{fig:out}, which shows the pulse-resolved azimuthally integrated intensity $I(q)$ before and after outlier pixel removal, along with the distribution of normalized RMS values. The results highlight a clear improvement in data quality following this procedure.

\subsection{Intensity filtering}
\begin{figure}
    \begin{center}
        \includegraphics[width=0.80\textwidth]{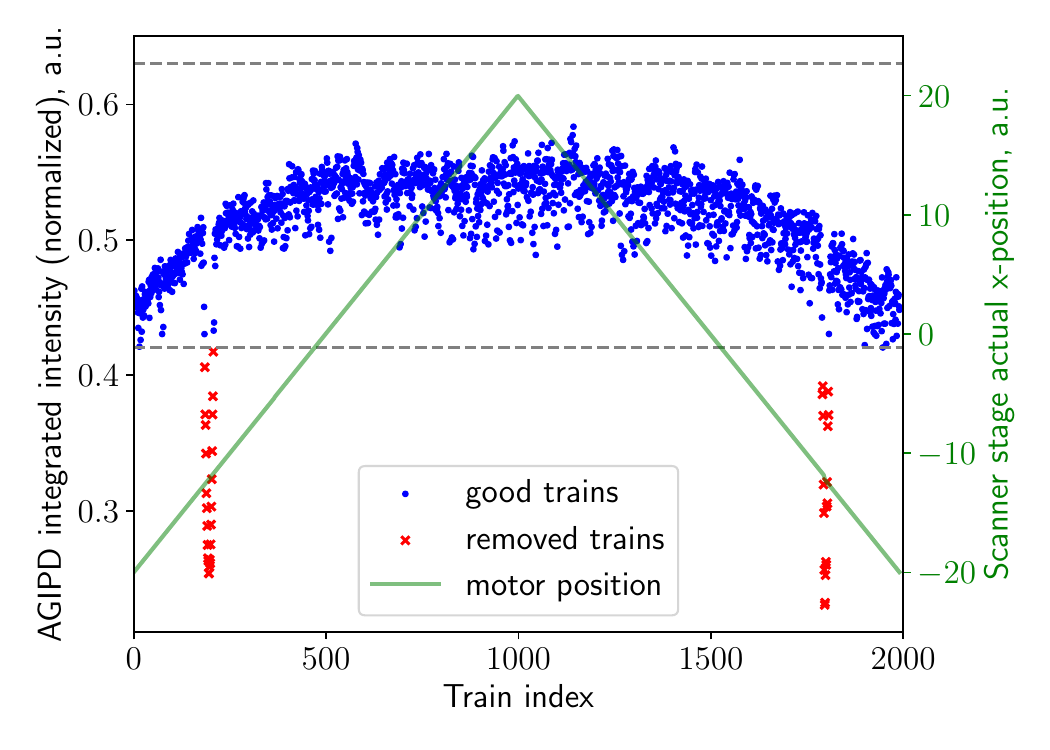}
        \caption{Example of applying intensity filtering to AGIPD data containing a signal drop due to an air bubble (red points) inside the capillary. The decrease in scattered intensity was encountered twice due to the back-and-forth scanning motion of the capillary, see Fig.~\ref{fig:stage_holder}c.}
        \label{fig:intens_filter}
    \end{center}
\end{figure}

In the experiment sample capillaries can be filled partially, have air bubbles or other inhomogeneities. This leads to SAXS intensity fluctuations in the data. To address this, the analysis pipeline includes an automatic filter for detecting poorly filled or missing sample based on changes in scattering intensity. The mean intensity values calculated in Section~\ref{ch3:mpc} are first normalized by the train-averaged X-ray gas monitor (XGM) signal \cite{Maltezopoulos2019}, accounting for fluctuations in the incoming X-ray intensity. All illuminated positions where the intensities (trains) are deviating by more than $\pm 20\%$ (this value is adjustable) from the median intensity value are excluded from the subsequent XPCS analysis, see Fig.~\ref{fig:intens_filter}. This filtering step must be applied with care to ensure removal of truly compromised  sample regions only, without discarding sample regions with intrinsic variations, {\em e.g.} caused by varying optical thicknesses.

\section{XPCS calculations}

A general approach to quantify the dynamics of both equilibrium and out-of-equilibrium systems is via the two-time correlation function TTCF \cite{sutton2002, madsen2010}

\begin{equation}
    TTCF (q, \tau_1, \tau_2) = \frac{\bigl\langle I(q,\tau_1)I(q,\tau_2) \bigr\rangle_q}{\bigl\langle I(q,\tau_1)\bigr\rangle_q \bigl\langle I(q,\tau_2)\bigr\rangle_q},
\label{ttcf_gen}
\end{equation}

\noindent which is defined as the correlation of intensity at two time points $\tau_1$ and $\tau_2$ for a set of detector pixels belonging to the same absolute scattering vector $q$. $\bigl\langle \cdots \bigr\rangle_q$ denotes averaging over the pixels from this set. For systems in equilibrium, the dynamics only depends on the time lag  $\Delta \tau = \tau_2 - \tau_1$ as \cite{madsen2010}
reducing the TTCF to the standard autocorrelation function:
\begin{equation}
g_2(q, \Delta \tau) = \bigl\langle TTCF(q,\tau_1, \Delta \tau) \bigr\rangle_{\tau_1}.
\end{equation}

A reduction in computational time and resources can be achieved by applying the multi-tau algorithm \cite{schatzel1990, cipelletti1999}, which is very efficient at synchrotron radiation facilities \cite{khan2018}, where XPCS is typically conducted in a "continuous" mode with uniform time intervals $\Delta \tau$ between successive frames.

However, this approach is not readily applicable to burst mode XPCS at EuXFEL, where the recorded frames are grouped within trains. In this mode, each frame's acquisition time is defined by the train-ID ($t_n$) and pulse-ID ($p_n$). This makes use of the multi-tau algorithm much less effective. To allow for the detection of potential beam-induced effects on sample dynamics, explicit calculation of the TTCFs is therefore preferred as given by

\begin{equation}
    TTCF (q,t_{n_1},p_{n_1},t_{n_2},p_{n_2}) = \frac{\bigl\langle I(q,t_{n_1},p_{n_1})I(q,t_{n_2},p_{n_2})\bigr\rangle_q}{\bigl\langle I(q,t_{n_1},p_{n_1})\bigr\rangle_q \bigl\langle I(q,t_{n_2},p_{n_2})\bigr\rangle_q}.
\end{equation}

Since each X-ray pulse train illuminates a fresh spot on the sample, correlations only need to be computed between pulses within the same train ($t_{n_1} = t_{n_2}$). This results in $n_{tr}$ independent TTCFs, each of dimension $(n_p, n_p)$, with $n_{tr}$ being the number of trains and $n_p$ the number of pulses per train. The final train-averaged TTCF is obtained as

\begin{equation}
    TTCF(q,p_{n_1},p_{n_2}) =  \bigl\langle TTCF(q,t_{n_1},p_{n_1},t_{n_2},p_{n_2}) \bigr\rangle_{t_{n_1} = t_{n_2}}.
    \label{eq:ttcf}
\end{equation}

To enable further correction procedures, correlations between neighboring trains $t_{n_2} = t_{n_1 + 1}$ (termed off-correlation) are also calculated. 

The train-wise computation of TTCFs as defined in Eq.~\ref{eq:ttcf} presents a significant computational challenge due to the large volume of experimental data involved. This results in substantial demands on both CPU resources and memory usage. To address this, the calculation is divided into several discrete steps, each optimized for parallel execution and resource efficiency: first, the $q$-bins of interest are defined ({\em i.e.}, pixel ROI-s in the form of annuli around the direct beam position), which later will be separated in the mathematical kernel to calculate TTCFs for each bin. These $q$-bins can either be specified by the user (for example, based on the data from the mean AGIPD image or azimuthally integrated intensity $I(q)$), or automatically generated by the pipeline as equidistant $q$-bins fitting inside the available detector area). Each $q$-bin is  represented by a boolean mask with shape $(16,512,128)$, matching the AGIPD's native data format. These masks are subsequently merged with the static mask, the calibration mask, and the  outlier mask as discussed above. 

In the second step, the TTCF calculation is performed independently for each AGIPD module/sequence file using multiprocessing. Each CPU core returns train-resolved data comprising: (i) the non-normalized TTCF ({\em i.e.} the numerator in (\ref{ttcf_gen}) prior to normalization by the average intensity values given by the denominator in (\ref{ttcf_gen})); (ii) the off-correlation between adjacent trains; (iii) the pulse-resolved total intensity (used for both TTCF and off-correlation data); and (iv) the number of pixels used for the TTCF / off-correlation calculation. A module/sequence resolved train-ID map is also generated to align entries across all AGIPD data files, which is important for multiprocessing and further merging of the data.

To avoid memory overflow during processing, we estimate the maximum memory load per core. For example, storing TTCFs and off-correlation data (volume equals the volume of TTCFs) for a single module/sequence file containing data for $n_{tr}=250$ trains, each with $n_p=352$ pulses and $n_{qbin}=15$ $q$-bins, results in a data volume of $n_{tr} \times n_{qbin} \times n_p \times n_p \times 4~\mathrm{bytes} \times 2 = 3.7~\mathrm{GB}$, where the factor $2$ accounts for storing both  TTCF and off-correlation arrays. Scaling this to all 175 module/sequence files yields approximately $176~\mathrm{files} \times 3.7~\mathrm{GiB} \approx 651~\mathrm{GB}$ of data, which exceeds the available memory on most nodes of the Maxwell cluster. To ensure that the  pipeline is executable with all nodes of the Maxwell cluster, we set a limit for the number of cores used in multiprocessing (flexible, based on the estimation above) and implement caching of data to disk after finishing the calculation for each individual module/sequence file. Guided by the performance benchmarks discussed in Sec.~\ref{ch3:mpc} (see also Fig.~\ref{fig:h5_read}), we set the default value of cores to be exploited during the calculation to $50$. If memory estimates exceeds the node capacity, this number is automatically reduced. This results in a memory usage of $\sim 185~\mathrm{GB}$, which is available on any node of the Maxwell cluster \cite{MaxwellHPC}.

\begin{figure}%[tb!]
	\includegraphics[width=0.80\linewidth]{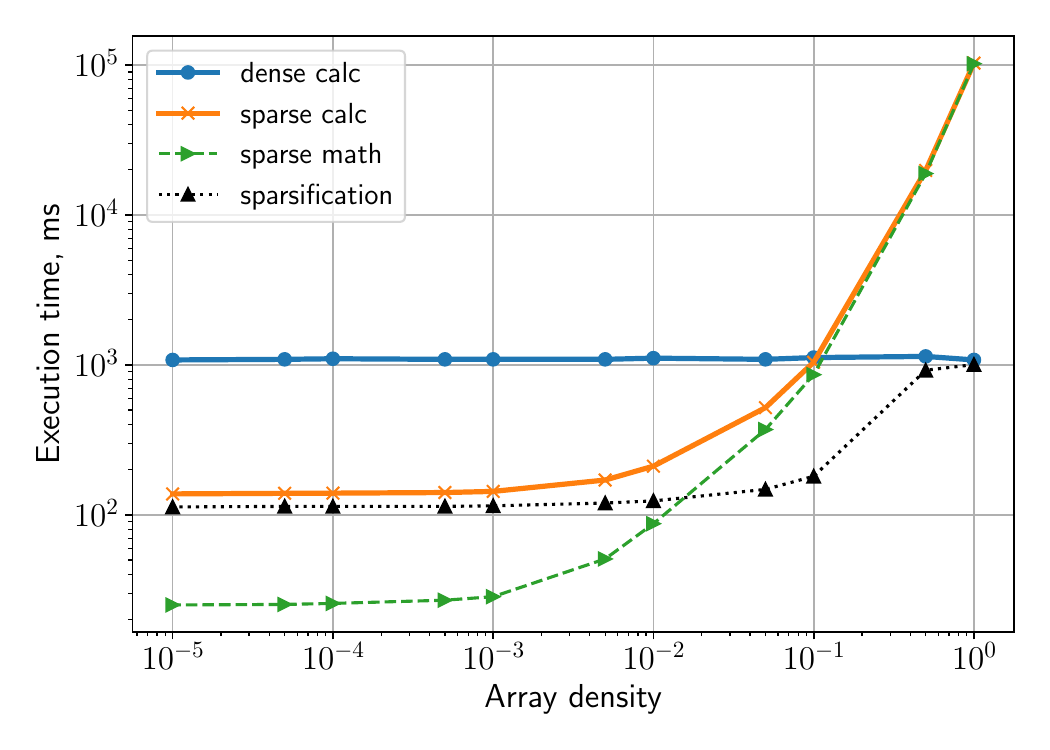}
	\caption{(Color online) Benchmarking of TTCF computation using an arbitrary data set with array shape $(300, 65536)$ and 10 $q$-bins, where $300$ is the number of time points and $65536=512 \times 128$ is the reshaped dimension of a single AGIPD module. The "dense calc" referes to standard matrix multiplication using dense arrays. The "sparse calc" includes both the time required to convert the data to sparse format ("sparsification") and the subsequent multiplication of sparse matrices ("sparse math").}
	\label{fig:sparse_dense}
\end{figure}

Another important factor influencing the overall runtime of the calculation is the performance of individual CPU cores. This performance can be optimized by considering the data density of the arrays to be processed, {\em i.e.} the proportion of non-zero values in the photonized data array relative to the total number of entries. For arrays with low photon density (where most of the pixel/cell values zero) computational performance can be improved by converting dense arrays into sparse formats and performing built-in linear algebra operations on the sparsified representation of the data. To evaluate this approach, we performed benchmarking tests using the native AGIPD array dimensions; the results are shown in Fig.~\ref{fig:sparse_dense}. We find that sparsification offers a computational advantage only below a certain data density threshold. Above an array density value of   $\sim 10^{-1}$ we observed a
loss of performance compared to conventional TTCF calculations using dense arrays.  
Therefore, to ensure optimal performance, the mean image density of each AGIPD module is evaluated, and the data representation—dense or sparse—is selected accordingly for each module/sequence file. The default threshold value for data sparsification is set to $5\times10^{-2}$.

\section{Post-processing and averaging}
\label{sec:off_corr}

After all module/sequence files have been processed and the corresponding results stored in the cache directory, the data must be merged. Specifically, for each train, the non-normalized TTCFs and off-correlations are summed across all detector modules, along with the corresponding total signal intensities and the number of contributing pixels. The train-ID map described earlier ensures synchronization of data across module/sequence files—that is, each chunk of data has a consistent length and train-ID ordering. This enables efficient use of linear algebra operations to merge data at the array level, avoiding slower element-wise (i.e., train-by-train) operations.

Subsequently, the TTCF and off-correlation matrices are normalized by the corresponding mean intensities, and the average photon count per pixel per pulse ($\bar{k}$) is computed for each train-ID/pulse-ID by dividing the mean intensity by the number of used pixels. To further improve data quality, the off-correlation matrix is subtracted from the TTCF to suppress residual artifacts from "jumping" pixels and other memory-cell related instabilties \cite{madsen2021, dallari2021analysis} (see Appendix for details). This correction also shifts the TTCF baseline from 1 to 0 as shown in Fig. \ref{fig:ttcf_off}. However, if the scattering intensity falls below $\bar{k}\lesssim 10^{-3}$ residual detector artifacts are still dominating the signal. Therefore, XPCS measurments at such low scattering intensities are considered impracticable with the AGIPD detector and the corresponding data entries are excluded from further analysis. Finally, the corrected and filtered data are averaged over all trains and the mean TTCFs and $g_2$ functions for each $q$-bin are stored for further analysis.

\section{Output}
\begin{figure}
    \begin{center}
        \includegraphics[width=1\textwidth]{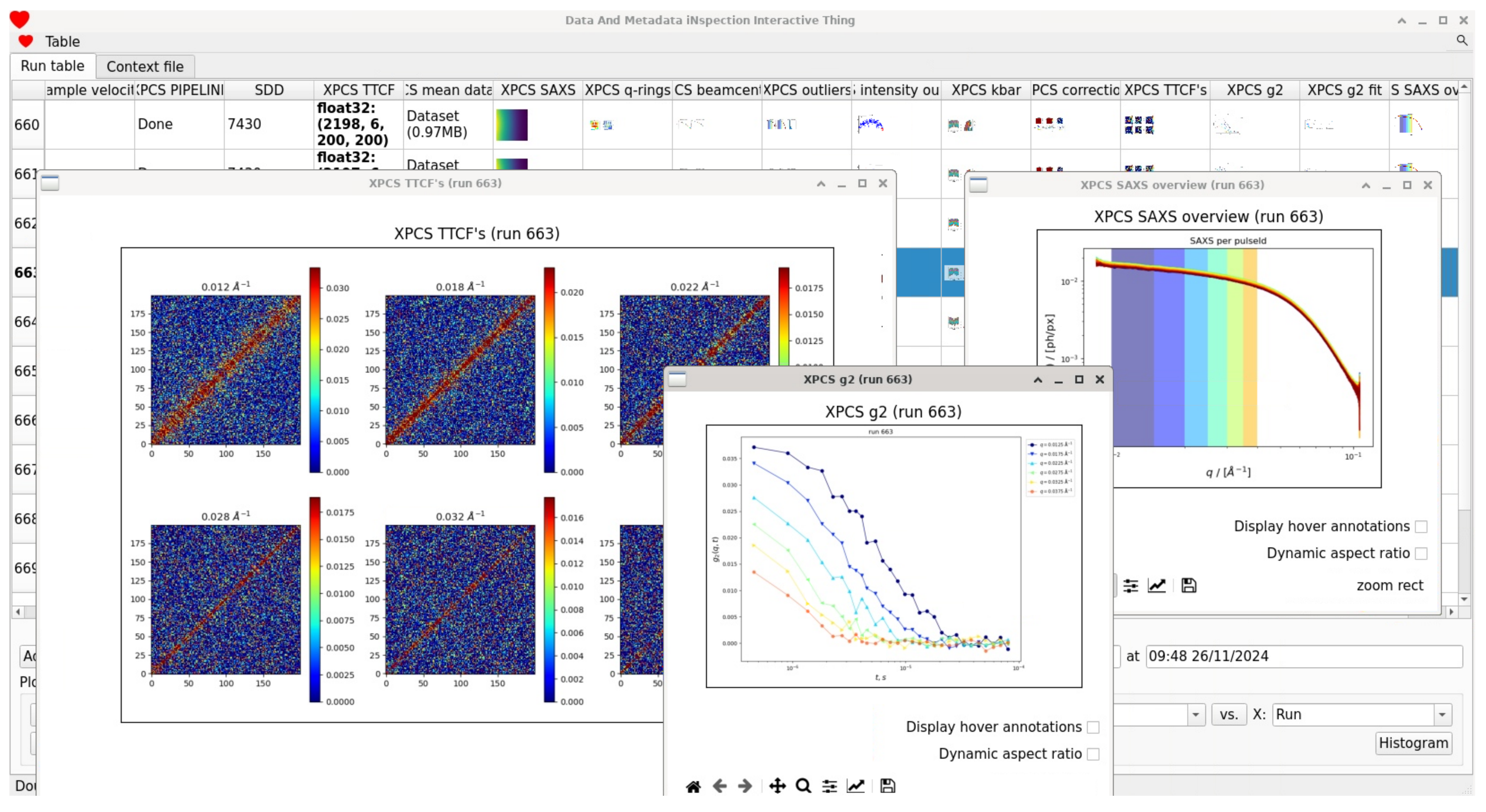}
        \caption{Example view of the DAMNIT tool showing a data table summarizing the results of the XPCS pipeline at the MID instrument. Each row corresponds to a single run, while columns display customizable metadata such as motor positions, monitor signals, and sample parameters.}
    \end{center}
    \label{fig:damnit}
\end{figure}

Users at the MID instrument have several options for inspecting and obtaining the output of the XPCS pipeline, inclduing the DAMNIT tool developed at EuXFEL \cite{damnit}. A graphical user interface (GUI) provides an overview of all runs acquired during a given experiment. A representative screenshot of the GUI is displayed in Fig. \ref{fig:damnit}. Each run is represented as a row in the overview table, while columns are customizable to display various types of data and metadata such as motor positions, monitor readings, or sample temperature. New runs are automatically added to the table and the corresponding fields are updated accordingly.

The output of the XPCS pipeline appears as columns in the DAMNIT table. Plots are represented by small icons, which can be opened as new figures. Data processed by the online XPCS pipeline, which is currently being developed and tested, is also transferred to DAMNIT. This allows the users to perform a preliminary analysis of XPCS data immediately after a run is finished, even before the fully calibrated detector data becomes available on the Maxwell cluster.

\begin{figure}
    \begin{center}
        \includegraphics[width=0.80\textwidth]{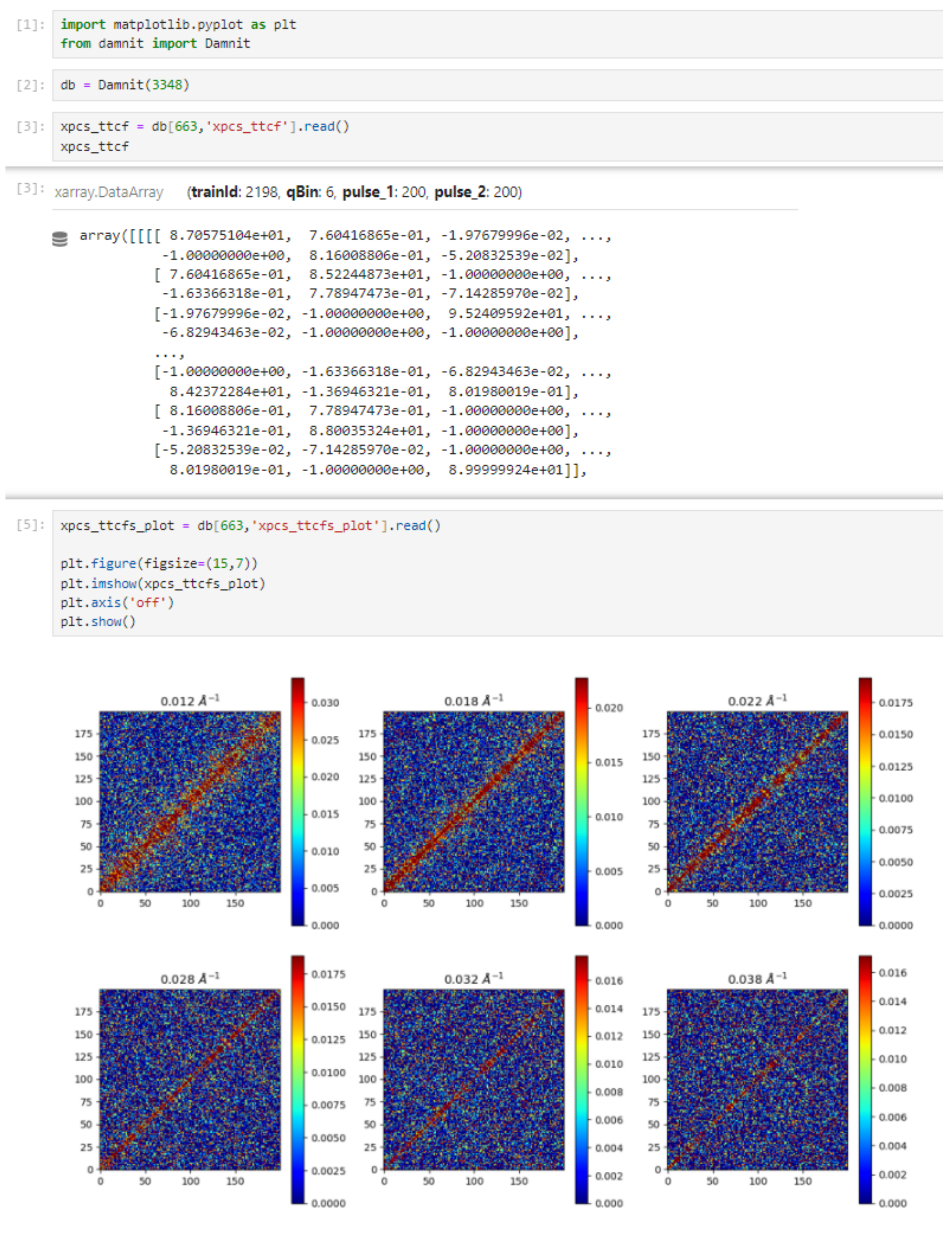}
        \caption{Accessing DAMNIT data through its Python API in a Jupyter notebook. The API allows users to retrieve, filter, and analyze experimental metadata and processed XPCS results directly within an interactive Python environment.}
    \end{center}
    \label{fig:jupyter}
\end{figure}

Further inspection and analysis of the XPCS results are enabled through Jupyter notebooks \cite{fangohr2019}, provided via the Jupyter Hub portal on the Maxwell cluster \cite{MaxJHUB}. Data contained in the DAMNIT table and stored in a database using HDF5 files, can be accessed using a dedicated Python API \cite{damnit}. An example usage is shown in Fig. \ref{fig:jupyter}, where users retrieve processed data arrays for further analysis using custom or external software. In addition to database access, output data from the XPCS pipeline is currently stored in the experiment directory as both image files and NetCDF4 format data. A future transition to NeXus-format output is planned to improve compatibility with community standards.

In line with the goals of the DAPHNE4NFDI consortium \cite{Barty_daphne, Murphy:ti5033}, efforts are ongoing to make MHz-XPCS data and metadata FAIR (Findable, Accessible, Interoperable, and Reusable). Current work focuses on establishing a secure connection between DAMNIT and the external DAPHNE4NFDI database (XPCS platform), which is envisioned as the central repository for high-quality processed and merged XPCS data. Following an initial embargo period, the platform is expected to provide open access to the broader scientific community.

\section{Discussion and conclusion}
\label{sec:future}

The MHz-XPCS data analysis pipeline presented here is designed to deliver results within a reasonable timeframe during the experiment, emphasizing minimal user interaction and high data quality suitable for publication. Based on operational experience made, several insights and suggestions have emerged that point towards desirable developments in detector capabilities and operation modes.

A reoccurring challenge in high-repetition-rate experiments using two-dimensional pixel detectors is the generation of high data volumes. These set significant demands on data transfer, storage, and computing infrastructure. The discrepancy between the data volumes read-out from the detector and the actual input to the final calculations can be staggering for sparse data techniques like single particle imaging or MHz-XPCS presented here.

Typical MHz-XPCS experiments yield fewer than one photon per pixel per frame, meaning the vast majority of pixels carry no intensity information. Nonetheless, full dense data arrays—including these zero-signal pixels—are repeatedly transferred from the detector to the online cluster and subsequently to the offline computing environment before any significant data reduction can occur (e.g., after dark pedestal subtraction). This inefficiency delays the identification and exclusion of empty pixels and represents a missed opportunity for early data compression.

To address this, implementing zero-value suppression as early as possible in the data acquisition chain—ideally on the detector itself—would significantly improve efficiency. Such functionality could be enabled by integrating real-time processing and filtering at the detector level using Field Programmable Gate Arrays (FPGAs). This concept of a “smart detector” capable of onboard data correction and selective readout is especially promising within the burst-mode operation scheme of the European XFEL. In the long term, this approach could offer a path toward more sustainable data handling and scalable analysis workflows.

It is also evident that high detector noise level complicates  the discrimination of single photons (single photon sensitivity), which results in artifacts like "jumping" pixels becoming problematic and also requiring more processing time ($\approx \times 2$ in both I/O and correlation operations due to the need for outlier detection and off-correlation). From a data reduction standpoint, high noise levels also hinder efficient compression. The conversion of analog signals to integer photon counts (“photonization”) becomes unreliable when photon signals cannot be clearly distinguished from noise. As a result, compression is no longer lossless, and residual uncertainties propagate through the analysis. Consequently, reducing detector noise not only enhances scientific data quality but also lowers computational demands and facilitates more efficient data reduction. Future detector developments that prioritize low-noise operation would enable both higher data quality and more efficient data processing.

Finally, we demonstrated that XPCS calculations in the sparse data regime benefit significantly from performing linear algebra operations on sparsified data representations.  However, the step of sparsification (converting the data from a dense array to a sparse array / list of events) can be more time consuming than the actual TTCF calculations. This makes detector concepts like the timepix4 development \cite{timepix4, Correa:yi5156}, which features an option of event based read-out (apart from traditional frame mode), very promising for future XPCS experiments, since a sparsified list representation of photon hits is directly output by the detector.
This not only streamlines processing but also opens the door to novel experimental regimes, all without requiring proportional increases in computational resources.

MHz-XPCS experiments are often constrained by the large pixel sizes of current detectors, which limit spatial resolution and reduce speckle contrast, thereby lowering the SNR of the correlation functions. 
This poses a particular challenge for radiation-sensitive samples, where increasing the beam size to reduce dose and dose rate effects is typically not an option. Additionally, extending XPCS into the wide-angle X-ray scattering (WAXS) regime—essential for accessing atomic length scales—is difficult with detectors like the AGIPD, due to the inherently lower speckle contrast at high scattering vectors \cite{lehmkuhler2021}. A factor 4 reduction in pixel size (e.g. from 200 $\mu$m to 50 $\mu$m) would improve the contrast and SNR significantly, but also increase the number of pixels by a factor of 16 (keeping the detector area constant). Still, the amount of data generated by an event-based detector would not scale proportionally, since the number of registered photons remains constant. Therefore, new detector developments in this direction hold the promise of expanding the range of experiments that are feasible, but also reducing the demands on data transfer, storage, and computation hardware to a fraction of the current requirements.

\section{Acknowledgements}
We acknowledge European XFEL in Schenefeld, Germany, for provision of X-ray free-electron laser beamtime at the Materials Imaging and Dynamics (MID) instrument at  SASE2 under proposal number 3118 and 3348, and would like to thank the staff for their assistance. Data recorded for the experiment at the European XFEL are available at doi:10.22003/XFEL.EU-DATA-003118-00 and doi:10.22003/XFEL.EU-DATA-003348-00. We acknowledge financial support by the consortium DAPHNE4NFDI in association with the German National Research Data Infrastructure (NFDI) e.V. - project number 4602487. This work was supported through the Maxwell computational resources operated at Deutsches Elektronen-Synchrotron DESY, Hamburg, Germany. We acknowledge J.~Sztuk-Dambietz, N.~Raab, O.~Meyer, T.~Laurus for their support, commissioning and fruitful comments on AGIPD operation. We acknowledge M.~Reiser, F.~Dallari and A.~Jain for the fruitful discussions on MHz-XPCS data analysis using AGIPD.

\referencelist[iucr.bbl]

\appendix
\section{"Jumping" pixels}

\begin{figure}
    \begin{center}
        \includegraphics[width=0.99\textwidth]{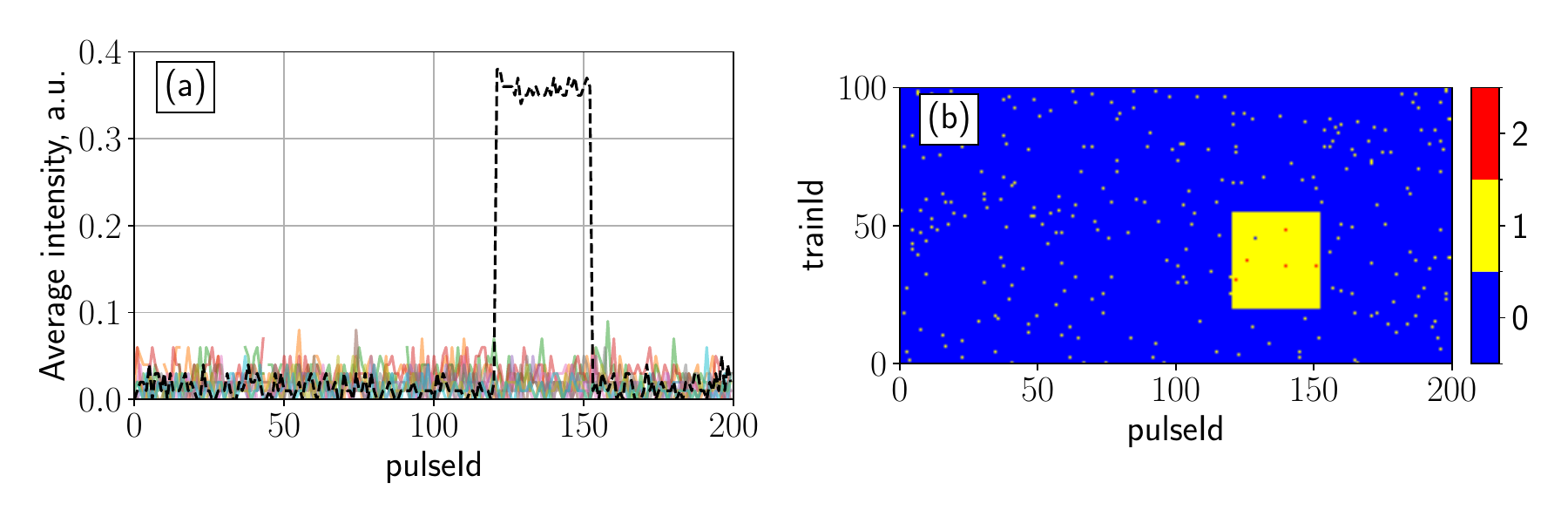}
        \caption{(a) Train-averaged pulse-resolved intensity of pixels within a selected q-bin: solid lines represent normal pixel behavior, whereas the dashed line represents typical behavior of the ''jumping pixel''; (b) Photon counts from the ''jumping'' pixel shown on the left, as a function of pulseId and trainId.}
        \label{fig:jump_pix}
    \end{center}
\end{figure}

\begin{figure}
    \begin{center}
        \includegraphics[width=0.99\textwidth]{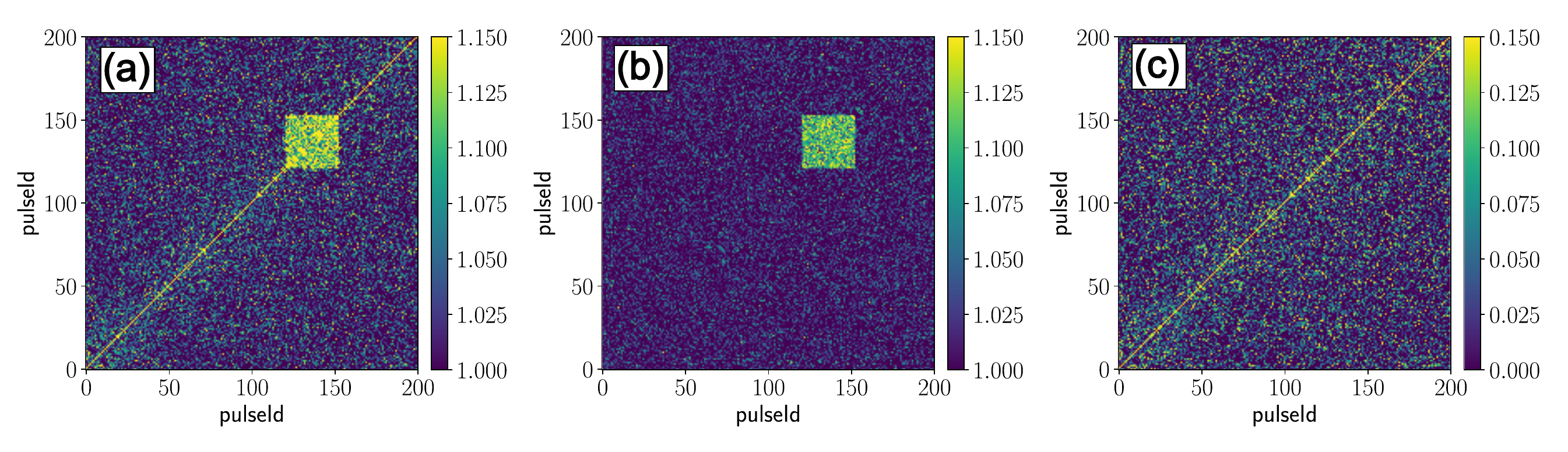}
        \caption{Performace of the off-correlation correction for a selected q-bin, containing the ''jumping'' pixel shown in Fig.\ref{fig:jump_pix}: (a) train-averaged TTCF without correction; (b) train-averaged off-correlation; (c) train-averaged TTCF after subtraction of the off-correlation (with this step the baseline shifts to 0).}
        \label{fig:ttcf_off}
    \end{center}
\end{figure}

''Jumping'' pixel is a phenomenon that describes the spontaneous drift of the pedestal of 32 consecutive memory cells of a single AGIPD pixel. Jumps can be observed on various time scales (from some trains to several minutes) and amplitudes (below and above a single photon equivalent), but always for one row of the $11 \times 32 = 352$ storage cell memory matrix. Most of the jumps are corrected by the common-mode correction or masked by the outlier detection. However, if the drift has an amplitude close to or above a single-photon signal and lasts only for a few consecutive trains, it might occasionally be missed. This can lead to clear false signal as shown in Fig.\ref{fig:jump_pix}, but might also result in more subtle higher photon count probabilities. The presence of such jumps results in $32\times32$ square artifacts in the TTCFs along the main diagonal. Fig.\ref{fig:ttcf_off}a shows an example of such an artifact arising from the presence of just a single ''jumping'' pixel from Fig.\ref{fig:jump_pix} within the selected q-bin.

Calculating the pulse-pulse correlations between the adjacent trains (off-correlations) can help to remove such artifacts as shown in Fig.\ref{fig:ttcf_off}b. Subtracting the off-correlation from the initial TTCF can significantly improve the TTCF behavior and remove the existing artifacts as shown in Fig.\ref{fig:ttcf_off}c. 

\end{document}